\documentclass{iaus}
\usepackage{graphicx}

\title[Atomic Data for Neutron-Capture Elements] 
{Advances in Atomic Data for Neutron-Capture Elements}

\author[N.\ C.\ Sterling, et al.]   
{N.\ C.\ Sterling$^1$
\thanks{NSF Astronomy and Astrophysics Postdoctoral Fellow},
M.\ C.\ Witthoeft$^{2,3}$.
D.\ A.\ Esteves$^4$,
P.\ C.\ Stancil$^5$
A.\ L.\ D.\ Kilcoyne$^6$,
R.\ C.\ Bilodeau$^{6,7}$,
\and A.\ Aguilar$^6$}

\affiliation{$^1$Department of Physics and Astronomy, Michigan State University,  3248 Biomedical Physical Sciences,
East Lansing, MI 38824-2320, USA, email: {\tt sterling@pa.msu.edu} \\[\affilskip]
$^2$NASA Goddard Space Flight Center, Code 662, Greenbelt, MD 20771, USA \\[\affilskip]
$^3$ Department of Astronomy, University of Maryland, College Park, MD 20742, USA \\[\affilskip]
$^4$ JILA, University of Colorado, Boulder, CO 80309-0440, USA \\[\affilskip]
$^5$ Department of Physics and Astronomy and the Center for Simulational Physics, University of Georgia, Athens, GA 30602-2451, USA\\[\affilskip]
$^6$ The Advanced Light Source, Lawrence Berkeley National Laboratory, Berkeley, CA 94720, USA \\[\affilskip]
$^7$ Western Michigan University, MS 5252, 1903 W.\ Michigan Ave., Kalamazoo, MI 49008, USA}

\pubyear{2012}
\volume{283}  
\jname{Planetary Nebulae: An Eye to the Future}
\begin{document}

\maketitle

\begin{abstract}
Neutron(\emph{n})-capture elements (atomic number $Z>30$), which can be produced in planetary nebula (PN) progenitor stars via \emph{s}-process nucleosynthesis, have been detected in nearly 100 PNe.  This demonstrates that nebular spectroscopy is a potentially powerful tool for studying the production and chemical evolution of trans-iron elements.  However, significant challenges must be addressed before this goal can be achieved.  One of the most substantial hurdles is the lack of atomic data for \emph{n}-capture elements, particularly that needed to solve for their ionization equilibrium (and hence to convert ionic abundances to elemental abundances).  To address this need, we have computed photoionization cross sections and radiative and dielectronic recombination rate coefficients for the first six ions of Se and Kr.  The calculations were benchmarked against experimental photoionization cross section measurements.  In addition, we computed charge transfer (CT) rate coefficients for ions of six \emph{n}-capture elements.  These efforts will enable the accurate determination of nebular Se and Kr abundances, allowing robust investigations of \emph{s}-process enrichments in PNe.
\end{abstract}

Neutron(\emph{n})-capture elements can be produced in low- and intermediate-mass stars ($\sim$1--8~M$_{\odot}$), the progenitors of planetary nebulae (PNe), during the asymptotic giant branch (AGB) phase via slow \emph{n}-capture nucleosynthesis (the ``\emph{s}-process''; \cite[Busso et al.\ 1999]{busso99}).  Nebular spectroscopy is a promising new tool for investigating \emph{n}-capture nucleosynthesis and the chemical evolution of trans-iron elements, providing access to elements not detectable in AGB stars and to classes of stars (e.g., intermediate-mass stars, 4--8~M$_{\odot}$) whose photospheres are obscured by heavy mass-loss during the AGB and post-AGB stages of evolution.  The detection of \emph{n}-capture element emission lines in the optical and near-infrared spectra of nearly 100 PNe (\cite[Sharpee et al.\ 2007]{sharpee07}, \cite[Sterling \& Dinerstein 2008]{sterling08}) demonstrates the potential of nebular spectroscopy for the study of these species.

The lack of atomic data for processes governing the ionization balance of \emph{n}-capture elements presents the primary challenge in honing nebular spectroscopy into an effective tool for studying trans-iron elements.  These data are needed because typically only one or two ions of \emph{n}-capture elements have been detected in individual PNe, and unobserved ions must be corrected for to derive total elemental abundances.  These ionization correction factors are most robustly determined via photoionization modeling, provided that data for photoionization (PI), radiative recombination (RR), dielectronic recombination (DR), and charge transfer (CT) are available.  However, data for these processes are unknown for the vast majority of trans-iron element ions, preventing abundance determinations more accurate than factors of 2--3 (\cite[Sterling et al.\ 2007]{sterling07}, \cite[Sterling \& Dinerstein 2008]{sterling08}).

We have determined these atomic data for Se and Kr, the most widely-detected \emph{n}-capture elements in PNe.  Using the AUTOSTRUCTURE atomic structure code (\cite[Badnell 2011]{badnell11}), we computed multi-configuration distorted-wave PI cross sections and RR and DR rate coefficients for the first six Se and Kr ions (\cite[Sterling \& Witthoeft 2011]{sterling11c}, \cite[Sterling 2011]{sterling11d}).  In nearly all cases, DR is the dominant recombination mechanism, with rate coefficients at 10$^4$~K exceeding those of RR by as much as 1--2 orders of magnitude.

Our calculations were benchmarked against experimental absolute PI cross sections measured at the Advanced Light Source synchrotron radiation facility at Lawrence Berkeley National Laboratory in California (\cite[Sterling et al.\ 2011]{sterling11b}, \cite[Esteves et al.\ 2011a,b]{esteves11a, esteves11b}).  These measurements were conducted with the merged-beams method on the Ion Photon Beams apparatus, with typical accuracies of 20--30\%.

Based on comparison to experimental measurements and the sensitivity of our results to \linebreak orbital radial scaling parameters, we estimate the direct PI cross sections to have uncertainties of 30--50\% for most Se and Kr ions.  RR rate coefficients are uncertain by $\leq$10\%, while those of DR have larger uncertainties ranging from a factor of two up to two orders of magnitude, due to the unknown energies of near-threshold autoionizing resonances.

CT rate coefficients were determined for the first five ions of several \emph{n}-capture elements that have been detected in PN spectra (Ge, Se, Br, Kr, Rb, and Xe), using the Landau-Zener and Demkov approximations (\cite[Sterling \& Stancil 2011]{sterling11e}).  These approximations generally are accurate to within a factor of three for transitions with large rate coefficients, and an order of magnitude for weaker ones (\cite[Butler \& Dalgarno 1980]{bd80}).

We are expanding the atomic database of the photoionization code Cloudy (\cite[Ferland et al.\ 1998]{ferland98}) up to Kr, using the newly determined atomic data.  We will use Cloudy to compute a grid of models for deriving robust ionization corrections for unobserved Se and Kr ions, enabling much more accurate abundance determinations than currently possible.  Moreover, we will test the sensitivity of abundance determinations to uncertainties in the atomic data, illuminating the atomic processes and species that require further analysis.  Such tests are illustrative of the abundance uncertainties of lighter elements with atomic data derived in a similar manner (especially iron-peak elements).

\end{document}